\documentclass[12pt,a4paper]{article}
\usepackage{amssymb}
\usepackage[margin=2cm]{geometry}

\begin{document}


%
\begin{titlepage}
\begin{center}

\hfill DFPD/2015/TH/19


\vskip 2cm

{\huge{\bf Freudenthal Duality in Gravity:\\ \vspace{3pt} from Groups of Type $E_7$ \\ \vspace{7pt} to Pre-Homogeneous Spaces}}

\vskip 0.5cm

{\large{\bf Alessio Marrani\,$^{1,2}$}}

\vskip 20pt

{\em $^1$ Centro Studi e Ricerche ``Enrico Fermi'',\\
Via Panisperna 89A, I-00184, Roma, Italy \vskip 5pt }

\vskip 10pt

{\em $^2$ Dipartimento di Fisica e Astronomia ``Galileo Galilei'', \\Universit\`a di Padova,\\ Via Marzolo 8, I-35131 Padova, Italy \vskip 5pt }

{email: {\tt Alessio.Marrani@pd.infn.it}} \\

    \vspace{10pt}

\end{center}

\vskip 2.2cm

\begin{center}
{\bf ABSTRACT}\\[3ex]
\end{center}
Freudenthal duality can be defined as an anti-involutive, non-linear map
acting on symplectic spaces. It was introduced in four-dimensional
Maxwell-Einstein theories coupled to a non-linear sigma model of scalar
fields.

In this short review, I will consider its relation to the $U$-duality Lie groups of type $E_{7}$
in extended supergravity theories, and comment on the relation between the
Hessian of the black hole entropy and the pseudo-Euclidean, rigid special
(pseudo)K\"{a}hler metric of the pre-homogeneous spaces associated to the $U$%
-orbits.

\vskip 2.2cm


\vskip 1.0cm

\begin{center}

Talk presented at the Conference\\ {\em Group Theory, Probability, and the
Structure of Spacetime} in honor of V.S.Varadarajan,\\ UCLA Mathematics
Department, Los Angeles, CA USA, November 7--9, 2014. \\To appear in a special issue of {\em ``p-Adic Numbers, Ultrametric
Analysis and Applications''}.

\end{center}

\end{titlepage}


\newpage

\section{Freudenthal Duality}

We start and consider the following Lagrangian density in four dimensions (%
\textit{cfr. e.g.} \cite{BGM}):%
\begin{equation}
\mathcal{L}=-\frac{R}{2}+\frac{1}{2}g_{ij}\left( \varphi \right) \partial
_{\mu }\varphi ^{i}\partial ^{\mu }\varphi ^{j}+\frac{1}{4}I_{\Lambda \Sigma
}\left( \varphi \right) F_{\mu \nu }^{\Lambda }F^{\Sigma \mid \mu \nu }+%
\frac{1}{8\sqrt{-G}}R_{\Lambda \Sigma }\left( \varphi \right) \epsilon ^{\mu
\nu \rho \sigma }F_{\mu \nu }^{\Lambda }F_{\rho \sigma }^{\Sigma },
\label{1}
\end{equation}%
describing Einstein gravity coupled to Maxwell (Abelian) vector fields and
to a non-linear sigma model of scalar fields (with no potential); note that $%
\mathcal{L}$ may -but does not necessarily need to - be conceived as the
bosonic sector of $D=4$ (\textit{ungauged}) supergravity theory. Out of the
Abelian two-form field strengths $F^{\Lambda }$'s, one can define their
duals $G_{\Lambda }$, and construct a symplectic vector :%
\begin{equation}
H:=\left( F^{\Lambda },G_{\Lambda }\right) ^{T},~~~^{\ast }G_{\Lambda \mid
\mu \nu }:=2\frac{\delta \mathcal{L}}{\delta F^{\Lambda \mid \mu \nu }}.
\end{equation}

We then consider the simplest solution of the equations of motion deriving
from $\mathcal{L}$, namely a static, spherically symmetric, asymptotically
flat, dyonic extremal black hole with metric \cite{Pap}%
\begin{equation}
ds^{2}=-e^{2U(\tau )}dt^{2}+e^{-2U(\tau )}\left[ \frac{d\tau ^{2}}{\tau ^{4}}%
+\frac{1}{\tau ^{2}}\left( d\theta ^{2}+\sin \theta d\psi ^{2}\right) \right]
,  \label{2}
\end{equation}%
where $\tau :=-1/r$. Thus, the two-form field strengths and their duals can
be fluxed on the two-sphere at infinity $S_{\infty }^{2}$ in such a
background, respectively yielding the electric and magnetic charges of the
black hole itself, which can be arranged in a symplectic vector $\mathcal{Q}$
:%
\begin{eqnarray}
p^{\Lambda } &:&=\frac{1}{4\pi }\int_{S_{\infty }^{2}}F^{\Lambda
},~~~q_{\Lambda }:=\frac{1}{4\pi }\int_{S_{\infty }^{2}}G_{\Lambda }, \\
\mathcal{Q} &:&=\left( p^{\Lambda },q_{\Lambda }\right) ^{T}.
\end{eqnarray}

Then, by exploiting the symmetries of the background (\ref{2}), the
Lagrangian (\ref{1}) can be dimensionally reduced from $D=4$ to $D=1$,
obtaining a 1-dimensional effective Lagrangian ($^{\prime }:=d/d\tau $) \cite%
{FGK}:%
\begin{equation}
\mathcal{L}_{D=1}=\left( U^{\prime }\right) ^{2}+g_{ij}\left( \varphi
\right) \varphi ^{i\prime }\varphi ^{j\prime }+e^{2U}V_{BH}\left( \varphi ,%
\mathcal{Q}\right) \label{v2-1}
\end{equation}%
along with the Hamiltonian constraint \cite{FGK}%
\begin{equation}
\left( U^{\prime }\right) ^{2}+g_{ij}\left( \varphi \right) \varphi
^{i\prime }\varphi ^{j\prime }-e^{2U}V_{BH}\left( \varphi ,\mathcal{Q}%
\right) =0.\label{v2-2}
\end{equation}%
The so-called \textquotedblleft effective black hole potential" $V_{BH}$
appearing in (\ref{v2-1}) and (\ref{v2-2}) is defined as \cite{FGK}%
\begin{equation}
V_{BH}\left( \varphi ,\mathcal{Q}\right) :=-\frac{1}{2}\mathcal{Q}^{T}%
\mathcal{M}\left( \varphi \right) \mathcal{Q},  \label{3}
\end{equation}%
in terms of the symplectic and symmetric matrix \cite{BGM}
\begin{eqnarray}
\mathcal{M} &:&=\left(
\begin{array}{cc}
\mathbb{I} & -R \\
0 & \mathbb{I}%
\end{array}%
\right) \left(
\begin{array}{cc}
I & 0 \\
0 & I^{-1}%
\end{array}%
\right) \left(
\begin{array}{cc}
\mathbb{I} & 0 \\
-R & \mathbb{I}%
\end{array}%
\right) =\left(
\begin{array}{ccc}
I+RI^{-1}R & ~ & -RI^{-1} \\
~ & ~ & ~ \\
-I^{-1}R &  & I^{-1}%
\end{array}%
\right) , \\
\mathcal{M}^{T} &=&\mathcal{M};~~~\mathcal{M}\Omega \mathcal{M}=\Omega ,
\end{eqnarray}%
where $\mathbb{I}$ denotes the identity, and $R\left( \varphi \right) $ and $%
I\left( \varphi \right) $ are the scalar-dependent matrices occurring in (%
\ref{1}); moreover, $\Omega $ stands for the symplectic metric ($\Omega
^{2}=-\mathbb{I}$). Note that, regardless of the invertibility of $R\left(
\varphi \right) $ and as a consequence of the physical consistence of the
kinetic vector matrix $I\left( \varphi \right) $, $\mathcal{M}$ is
negative-definite; thus, the effective black hole potential (\ref{3}) is
positive-definite.

By virtue of the matrix $\mathcal{M}$, one can introduce a
(scalar-dependent) \textit{anti-involution} $\mathcal{S}$ in any
Maxwell-Einstein-scalar theory described by (\ref{1}) with a symplectic
structure $\Omega $, as follows :%
\begin{eqnarray}
\mathcal{S}\left( \varphi \right) &:&=\Omega \mathcal{M}\left( \varphi
\right) ;  \label{4} \\
\mathcal{S}^{2}\left( \varphi \right) &=&\Omega \mathcal{M}\left( \varphi
\right) \Omega \mathcal{M}\left( \varphi \right) =\Omega ^{2}=-\mathbb{I};
\end{eqnarray}%
in turn, this allows to define an anti-involution on the dyonic charge
vector $\mathcal{Q}$, which has been called (scalar-dependent) \textit{%
Freudenthal duality} \cite{FD-Duff,FDY,F-Dual-L}:%
\begin{eqnarray}
\mathfrak{F}\left( \mathcal{Q};\varphi \right) &:&=-\mathcal{S}\left(
\varphi \right) \mathcal{Q};  \label{FD} \\
\mathfrak{F}^{2} &=&-\mathbb{I},~~~(\forall \left\{ \varphi \right\} ).
\end{eqnarray}%
By recalling (\ref{3}) and (\ref{4}), the action of $\mathfrak{F}$ on $%
\mathcal{Q}$, defining the so-called ($\varphi $-dependent) Freudenthal dual
of $\mathcal{Q}$ itself, can be related to the symplectic gradient of the
effective black hole potential $V_{BH}$ :
\begin{equation}
\mathfrak{F}\left( \mathcal{Q};\varphi \right) =\Omega \frac{\partial
V_{BH}\left( \varphi ,\mathcal{Q}\right) }{\partial \mathcal{Q}}.
\end{equation}

Through the attractor mechanism \cite{AM-Refs}, all this enjoys an
interesting physical interpretation when evaluated at the (unique) event
horizon of the extremal black hole (\ref{2}) (denoted below by the subscript
\textquotedblleft $H$"); indeed%
\begin{eqnarray}
\partial _{\varphi }V_{BH} &=&0\Leftrightarrow \lim_{\tau \rightarrow
-\infty }\varphi ^{i}\left( \tau \right) =\varphi _{H}^{i}\left( \mathcal{Q}%
\right) ; \\
S_{BH}\left( \mathcal{Q}\right) &=&\frac{A_{H}}{4}=\pi \left.
V_{BH}\right\vert _{\partial _{\varphi }V_{BH}=0}=-\frac{\pi }{2}\mathcal{Q}%
^{T}\mathcal{M}_{H}\left( \mathcal{Q}\right) \mathcal{Q},
\end{eqnarray}%
where $S_{BH}$ and $A_{H}$ respectively denote the Bekenstein-Hawking
entropy \cite{BH} and the area of the horizon of the extremal black hole,
and the matrix horizon value $\mathcal{M}_{H}$ is defined as%
\begin{equation}
\mathcal{M}_{H}\left( \mathcal{Q}\right) :=\lim_{\tau \rightarrow -\infty }%
\mathcal{M}\left( \varphi \left( \tau \right) \right) .
\end{equation}%
Correspondingly, one can define the (scalar-independent) horizon Freudenthal
duality $\mathfrak{F}_{H}$ as the horizon limit of (\ref{FD}) :%
\begin{equation}
\widetilde{\mathcal{Q}}\equiv \mathfrak{F}_{H}\left( \mathcal{Q}\right)
:=\lim_{\tau \rightarrow -\infty }\mathfrak{F}\left( \mathcal{Q};\varphi
\left( \tau \right) \right) =-\Omega \mathcal{M}_{H}\left( \mathcal{Q}%
\right) \mathcal{Q}=\frac{1}{\pi }\Omega \frac{\partial S_{BH}\left(
\mathcal{Q}\right) }{\partial \mathcal{Q}}.
\end{equation}%
Remarkably, the (horizon) Freudenthal dual of $\mathcal{Q}$ is nothing but ($%
1/\pi $ times) the symplectic gradient of the Bekenstein-Hawking black hole
entropy $S_{BH}$; this latter, from dimensional considerations, is only
constrained to be an homogeneous function of degree two in $\mathcal{Q}$. As
a result, $\widetilde{\mathcal{Q}}=\widetilde{\mathcal{Q}}\left( \mathcal{Q}%
\right) $ is generally a complicated (non-linear) function, homogeneous of
degree one in $\mathcal{Q}$.

It can be proved that the entropy $S_{BH}$ itself is invariant along the
flow in the charge space $\mathcal{Q}$ defined by the symplectic gradient
(or, equivalently, by the horizon Freudenthal dual) of $\mathcal{Q}$ itself :%
\begin{equation}
S_{BH}\left( \mathcal{Q}\right) =S_{BH}\left( \mathfrak{F}_{H}\left(
\mathcal{Q}\right) \right) =S_{BH}\left( \frac{1}{\pi }\Omega \frac{\partial
S_{BH}\left( \mathcal{Q}\right) }{\partial \mathcal{Q}}\right) =S_{BH}\left(
\widetilde{\mathcal{Q}}\right) .  \label{inv-S}
\end{equation}

It is here worth pointing out that this invariance is pretty remarkable :
the (semi-classical) Bekenstein-Hawking entropy of an extremal black hole
turns out to be invariant under a generally non-linear map acting on the
black hole charges themselves, and corresponding to a symplectic gradient
flow in their corresponding vector space.

For other applications and instances of Freudenthal duality, see \cite%
{Ortin,FFM,Freudenthal-GT}.

\section{Groups of Type $E_{7}$}

The concept of Lie groups \textit{of type }$\mathit{E}_{7}$ as introduced in
the 60s by Brown \cite{Brown}, and then later developed \textit{e.g.} by
\cite{Meyberg,Garibaldi,Krutelevich,FKM,Exc-Reds}.

Starting from a pair $(G,\mathbf{R})$ made of a Lie group $G$ and its
faithful representation $\mathbf{R}$, the three axioms defining ($G,\mathbf{R%
}$) as a group of type $E_{7}$ read as follows :

\begin{enumerate}
\item Existence of a (unique) symplectic invariant structure $\Omega $ in $%
\mathbf{R}$ :%
\begin{equation}
\exists !\Omega \equiv \mathbf{1}\in \mathbf{R}\times _{a}\mathbf{R},
\end{equation}%
which then allows to define a symplectic product $\left\langle \cdot ,\cdot
\right\rangle $ among two vectors in the representation space $\mathbf{R}$
itself :%
\begin{equation}
\left\langle Q_{1},Q_{2}\right\rangle :=Q_{1}^{M}Q_{2}^{N}\Omega
_{MN}=-\left\langle Q_{2},Q_{1}\right\rangle .
\end{equation}

\item Existence of \ (unique) rank-4 completely symmetric invariant tensor ($%
K$-tensor) in $\mathbf{R}$ :%
\begin{equation}
\exists !K\equiv \mathbf{1}\in \left( \mathbf{R}\times \mathbf{R}\times
\mathbf{R}\times \mathbf{R}\right) _{s},
\end{equation}%
which then allows to define a degree-4 invariant polynomial $I_{4}$ in $%
\mathbf{R}$ itself :%
\begin{equation}
I_{4}:=K_{MNPQ}Q^{M}Q^{N}Q^{P}Q^{Q}.  \label{I4}
\end{equation}

\item Defining a triple map $T$ in $\mathbf{R}$ as%
\begin{eqnarray}
T &:&\mathbf{R\times R\times R\rightarrow R}; \\
\left\langle T\left( Q_{1},Q_{2},Q_{3}\right) ,Q_{4}\right\rangle
&:&=K_{MNPQ}Q_{1}^{M}Q_{2}^{N}Q_{3}^{P}Q_{4}^{Q},
\end{eqnarray}%
it holds that%
\begin{equation}
\left\langle T\left( Q_{1},Q_{1},Q_{2}\right) ,T\left(
Q_{2},Q_{2},Q_{2}\right) \right\rangle =\left\langle
Q_{1},Q_{2}\right\rangle K_{MNPQ}Q_{1}^{M}Q_{2}^{N}Q_{2}^{P}Q_{2}^{Q}.
\end{equation}%
This property makes a group of type $E_{7}$ amenable to a description as an
automorphism group of a \textit{Freudenthal triple system} (or,
equivalently, as the conformal groups of the underlying Jordan triple system
- whose a Jordan algebra is a particular case - ).
\end{enumerate}

All electric-magnetic duality ($U$-duality\footnote{%
Here $U$-duality is referred to as the \textquotedblleft
continuous\textquotedblright\ symmetries of \cite{CJ-1}. Their discrete
versions are the $U$-duality non-perturbative string theory symmetries
introduced by Hull and Townsend \cite{HT-1}.}) groups of $\mathcal{N}%
\geqslant 2$-extended $D=4$ supergravity theories with symmetric scalar
manifolds are of type $E_{7}$. Among these, degenerate groups of type $E_{7}$
are those in which the $K$-tensor is actually reducible, and thus $I_{4}$ is
the square of a quadratic invariant polynomial $I_{2}$. In fact, in general,
in theories with electric-magnetic duality groups of type $E_{7}$ holds that%
\begin{equation}
S_{BH}=\pi \sqrt{\left\vert I_{4}\left( \mathcal{Q}\right) \right\vert }=\pi
\sqrt{\left\vert K_{MNPQ}\mathcal{Q}^{M}\mathcal{Q}^{N}\mathcal{Q}^{P}%
\mathcal{Q}^{Q}\right\vert },
\end{equation}%
whereas in the case of degenerate groups of type $E_{7}$ it holds that $%
I_{4}\left( \mathcal{Q}\right) =\left( I_{2}\left( \mathcal{Q}\right)
\right) ^{2}$, and therefore the latter formula simplifies to%
\begin{equation}
S_{BH}=\pi \sqrt{\left\vert I_{4}\left( \mathcal{Q}\right) \right\vert }=\pi
\left\vert I_{2}\left( \mathcal{Q}\right) \right\vert .
\end{equation}

\begin{table}[h!]
\begin{center}
\begin{tabular}{|c||c|c|c|}
\hline
$%
\begin{array}{c}
\\
J_{3}%
\end{array}
$ & $%
\begin{array}{c}
\\
G_{4} \\
~~%
\end{array}
$ & $%
\begin{array}{c}
\\
\mathbf{R} \\
~~%
\end{array}
$ & $%
\begin{array}{c}
\\
\mathcal{N} \\
~~%
\end{array}
$ \\ \hline\hline
$%
\begin{array}{c}
\\
J_{3}^{\mathbb{O}} \\
~%
\end{array}
$ & $E_{7\left( -25\right) }~$ & $\mathbf{56}$ & $2~$ \\ \hline
$%
\begin{array}{c}
\\
J_{3}^{\mathbb{O}_{s}} \\
~%
\end{array}
$ & $E_{7\left( 7\right) }$ & $\mathbf{56}$ & $8$ \\ \hline
$%
\begin{array}{c}
\\
J_{3}^{\mathbb{H}} \\
~%
\end{array}
$ & $SO^{\ast }\left( 12\right) $ & $\mathbf{32}$ & $2,~6$ \\ \hline
$%
\begin{array}{c}
\\
J_{3}^{\mathbb{H}_{s}} \\
~%
\end{array}
$ & $SO\left( 6,6\right) $ & $\mathbf{32}~$ & $0$ \\ \hline
$%
\begin{array}{c}
\\
J_{3}^{\mathbb{C}} \\
~%
\end{array}
$ & $SU\left( 3,3\right) $ & $\mathbf{20}$ & $2~$ \\ \hline
$%
\begin{array}{c}
\\
J_{3}^{\mathbb{C}_{s}} \\
~%
\end{array}
$ & $SL\left( 6,\mathbb{R}\right) $ & $\mathbf{20}$ & $0$ \\ \hline
$%
\begin{array}{c}
\\
M_{1,2}\left( \mathbb{O}\right) \\
~%
\end{array}
$ & $SU\left( 1,5\right) $ & $\mathbf{20}$ & $5$ \\ \hline
$%
\begin{array}{c}
\\
J_{3}^{\mathbb{R}} \\
~%
\end{array}
$ & $Sp\left( 6,\mathbb{R}\right) $ & $\mathbf{14}^{\prime }$ & $2$ \\ \hline
$%
\begin{array}{c}
\\
\mathbb{R}\oplus \mathbb{R}\oplus \mathbb{R} \\
(STU)~%
\end{array}
$ & $\left[ SL\left( 2,\mathbb{R}\right) \right] ^{3}$ & $\left( \mathbf{2},%
\mathbf{2},\mathbf{2}\right) $ & $2$ \\ \hline
$%
\begin{array}{c}
\\
\mathbb{R} \\
(T^{3})~~%
\end{array}
$ & $SL\left( 2,\mathbb{R}\right) $ & $\mathbf{4}$ & $2$ \\ \hline
\end{tabular}%
\end{center}
\caption{Simple, non-degenerate groups $G$ related to Freudenthal triple
systems $\mathfrak{M}\left( J_{3}\right) $ on simple rank-$3$ Jordan
algebras $J_{3}$. In general, $G\protect\cong Conf\left( J_{3}\right)
\protect\cong Aut\left( \mathfrak{M}\left( J_{3}\right) \right) $ (see
\textit{e.g.} \protect\cite{G-Lects,Small-Orbits-Phys,Small-Orbits-Maths}
for a recent introduction, and a list of Refs.). $\mathbb{O}$, $\mathbb{H}$,
$\mathbb{C}$ and $\mathbb{R}$ respectively denote the four division algebras
of octonions, quaternions, complex and real numbers, and $\mathbb{O}_{s}$, $%
\mathbb{H}_{s}$, $\mathbb{C}_{s}$ are the corresponding split forms. Note
that the $G$ related to split forms $\mathbb{O}_{s}$, $\mathbb{H}_{s}$, $%
\mathbb{C}_{s}$ is the \textit{maximally non-compact} (\textit{split}) real
form of the corresponding compact Lie group. $M_{1,2}\left( \mathbb{O}%
\right) $ is the Jordan triple system generated by $2\times 1$ vectors over $%
\mathbb{O}$ \protect\cite{GST}. Note that the $STU$ model, based on $J_{3}=%
\mathbb{R}\oplus \mathbb{R}\oplus \mathbb{R}$, has a \textit{semi-simple} $%
G_{4}$, but its \textit{triality symmetry} \protect\cite{stu} renders it
``effectively simple''. The $D=5$ uplift of the $T^{3}$ model based on $%
J_{3}=\mathbb{R}$ is the \textit{pure} $\mathcal{N}=2$, $D=5$ supergravity. $%
J_{3}^{\mathbb{H}}$ is related to both $8$ and $24$ supersymmetries, because
the corresponding supergravity theories are \textit{\textquotedblleft twin"}%
, namely they share the very same bosonic sector \protect\cite%
{GST,ADF-fixed,Gnecchi-1,Samtleben-Twin}. }
\end{table}

Simple, non-degenerate groups of type $E_{7}$ relevant to $\mathcal{N}%
\geqslant 2$-extended $D=4$ supergravity theories with symmetric scalar
manifolds are reported in Table 1.

Semi-simple, non-degenerate groups of type $E_{7}$ of the same kind are
given by $G=SL(2,\mathbb{R})\times SO(2,n)$ and $G=SL(2,\mathbb{R})\times
SO(6,n)$, with $\mathbf{R}=\left( \mathbf{2},\mathbf{2+n}\right) $ and $%
\mathbf{R}=\left( \mathbf{2},\mathbf{6+n}\right) $, respectively relevant
for $\mathcal{N}=2$ and $\mathcal{N}=4$ supergravity.

Moreover, degenerate (simple) groups of type $E_{7}$ relevant to the same
class of theories are $G=U(1,n)$ and $G=U(3,n)$, with complex fundamental
representations $\mathbf{R}=\mathbf{n+1}$ and $\mathbf{R}=\mathbf{3+n}$,
respectively relevant for $\mathcal{N}=2$ and $\mathcal{N}=3$ supergravity
\cite{FKM}.

The classification of groups of type $E_{7}$ is still an open problem, even
if some progress have been recently made \textit{e.g.} in \cite{G2} (in
particular, \textit{cfr.} Table D therein).

In all the aforementioned cases, the scalar manifold is a \textit{symmetric}
cosets $\frac{G}{H}$, where $H$ is the maximal compact subgroup (with
symmetric embedding) of $G$. Moreover, the $K$-tensor can generally be
expressed as \cite{Exc-Reds}%
\begin{equation}
K_{MNPQ}=-\frac{n(2n+1)}{6d}\left[ t_{MN}^{\alpha }t_{\alpha \mid PQ}-\frac{d%
}{n\left( 2n+1\right) }\Omega _{M(P}\Omega _{Q)N}\right] ,
\end{equation}%
where $\dim \mathbf{R}=2n$ and $\dim G=d$, and $t_{MN}^{\alpha }$ denotes
the symplectic representation of the generators of $G$ itself. Thus, the
horizon Freudenthal duality can be expressed in terms of the $K$-tensor as
follows \cite{FD-Duff}:%
\begin{equation}
\mathfrak{F}_{H}\left( \mathcal{Q}\right) _{M}\equiv \widetilde{\mathcal{Q}}%
_{M}=\frac{\partial \sqrt{\left\vert I_{4}\left( \mathcal{Q}\right)
\right\vert }}{\partial \mathcal{Q}^{M}}=\epsilon \frac{2}{\sqrt{\left\vert
I_{4}\left( \mathcal{Q}\right) \right\vert }}K_{MNPQ}\mathcal{Q}^{N}\mathcal{%
Q}^{P}\mathcal{Q}^{Q},
\end{equation}%
where $\epsilon :=I_{4}/\left\vert I_{4}\right\vert $; note that the horizon
Freudenthal dual of a given symplectic dyonic charge vector $\mathcal{Q}$ is
well defined only when $\mathcal{Q}$ is such that $I_{4}\left( \mathcal{Q}%
\right) \neq 0$. Consequently, the invariance (\ref{inv-S}) of the black
hole entropy under the the horizon Freudenthal duality can be recast as the
invariance of $I_{4}$ itself :%
\begin{equation}
I_{4}\left( \mathcal{Q}\right) =I_{4}\left( \widetilde{\mathcal{Q}}\right)
=I_{4}\left( \Omega \frac{\partial \sqrt{\left\vert I_{4}\left( \mathcal{Q}%
\right) \right\vert }}{\partial \mathcal{Q}}\right) .
\end{equation}

In absence of \textquotedblleft flat directions" at the attractor points
(namely, of unstabilized scalar fields at the horizon of the black hole),
and for $I_{4}>0$, the expression of the matrix $\mathcal{M}_{H}\left(
\mathcal{Q}\right) $ at the horizon can be computed to read%
\begin{equation}
\mathcal{M}_{H\mid MN}(\mathcal{Q})=-\frac{1}{\sqrt{I_{4}}}\left( 2%
\widetilde{\mathcal{Q}}_{M}\widetilde{\mathcal{Q}}_{N}-6K_{MNPQ}\mathcal{Q}%
^{P}\mathcal{Q}^{Q}+\mathcal{Q}_{M}\mathcal{Q}_{N}\right) ,  \label{pom}
\end{equation}%
and it is invariant under horizon Freudenthal duality :%
\begin{equation}
\mathfrak{F}_{H}\left( \mathcal{M}_{H}\right) _{MN}:=\mathcal{M}_{H\mid MN}(%
\widetilde{\mathcal{Q}})=\mathcal{M}_{H\mid MN}(\mathcal{Q}).
\end{equation}

\section{Duality Orbits, Rigid Special K\"{a}hler Geometry and
Pre-Homogeneous Vector Spaces}

For $I_{4}>0$, $\mathcal{M}_{H}\left( \mathcal{Q}\right) $ given by (\ref%
{pom}) is one of the two possible solutions to the set of equations \cite%
{Dualities}%
\begin{equation}
\left\{
\begin{array}{l}
M^{T}\left( \mathcal{Q}\right) \Omega M\left( \mathcal{Q}\right) =\epsilon
\Omega ; \\
\\
M^{T}\left( \mathcal{Q}\right) =M\left( \mathcal{Q}\right) ; \\
\\
\mathcal{Q}^{T}M\left( \mathcal{Q}\right) \mathcal{Q}=-2\sqrt{\left\vert
I_{4}\left( \mathcal{Q}\right) \right\vert },%
\end{array}%
\right.  \label{sys}
\end{equation}%
which describes symmetric, purely $\mathcal{Q}$-dependent structures at the
horizon; they are symplectic or anti-symplectic, depending on whether $%
I_{4}>0$ or $I_{4}<0$, respectively. Since in the class of (super)gravity $%
D=4$ theories discussed the sign of $I_{4}$ actually determines a
stratification of the representation space $\mathbf{R}$ of charges into
distinct orbits of the action of $G$ into $\mathbf{R}$ itself (usually named
duality orbits), the symplectic or anti-symplectic nature of the solutions
to the system (\ref{sys}) is $G$-invariant, and supported by the various
duality orbits of $G$ (in particular, by the so-called \textquotedblleft
large" orbits, for which $I_{4}$ is non-vanishing).

One of the two possible solutions to the system (\ref{sys}) reads \cite%
{Dualities}%
\begin{eqnarray*}
M_{+}(\mathcal{Q}) &=&-\frac{1}{\sqrt{\left\vert I_{4}\right\vert }}\left( 2%
\widetilde{\mathcal{Q}}_{M}\widetilde{\mathcal{Q}}_{N}-6\epsilon K_{MNPQ}%
\mathcal{Q}^{P}\mathcal{Q}^{Q}+\epsilon \mathcal{Q}_{M}\mathcal{Q}%
_{N}\right) ; \\
\mathfrak{F}_{H}\left( M_{+}\right) _{MN} &:&=M_{+\mid MN}(\widetilde{%
\mathcal{Q}})=\epsilon M_{+\mid MN}(\mathcal{Q}).
\end{eqnarray*}%
For $\epsilon =+1\Leftrightarrow I_{4}>0$, it thus follows that%
\begin{equation}
M_{+}(\mathcal{Q})=\mathcal{M}_{H}\left( \mathcal{Q}\right) ,
\end{equation}%
as anticipated.

On the other hand, the other solution to system (\ref{sys}) reads \cite%
{Dualities}%
\begin{eqnarray}
M_{-}\left( \mathcal{Q}\right) &=&\frac{1}{\sqrt{\left\vert I_{4}\right\vert
}}\left( \widetilde{\mathcal{Q}}_{M}\widetilde{\mathcal{Q}}_{N}-6\epsilon
K_{MNPQ}\mathcal{Q}^{P}\mathcal{Q}^{Q}\right) ; \\
\mathfrak{F}_{H}\left( M_{-}\right) _{MN} &:&=M_{-\mid MN}(\widetilde{%
\mathcal{Q}})=\epsilon M_{-\mid MN}(\mathcal{Q}).
\end{eqnarray}%
By recalling the definition of $I_{4}$ (\ref{I4}), it is then immediate to
realize that $M_{-}\left( \mathcal{Q}\right) $ is the (opposite of the)
Hessian matrix of ($1/\pi $ times) the black hole entropy $S_{BH}$ :%
\begin{equation}
M_{-\mid MN}\left( \mathcal{Q}\right) =-\partial _{M}\partial _{N}\sqrt{%
\left\vert I_{4}\right\vert }=-\frac{1}{\pi }\partial _{M}\partial
_{N}S_{BH}.  \label{M-}
\end{equation}

The matrix $M_{-}\left( \mathcal{Q}\right) $ is the (opposite of the)
pseudo-Euclidean metric of a non-compact, non-Riemannian rigid special K\"{a}%
hler manifold related to the duality orbit of the black hole electromagnetic
charges (to which $\mathcal{Q}$ belongs), which is an example of
pre-homogeneous vector space (PVS) \cite{Sato}. In turn, the nature of the
rigid special manifold may be K\"{a}hler or pseudo-K\"{a}hler, depending on
the existence of a $U(1)$ or $SO(1,1)$ connection\footnote{%
For a thorough introduction to special K\"{a}hler geometry, see \textit{e.g.}
\cite{Freed}.}.

In order to clarify this statement, let us make two examples within maximal $%
\mathcal{N}=8$, $D=4$ supergravity. In this theory, the electric-magnetic
duality group is $G=E_{7(7)}$, and the representation in which the e.m.
charges sit is its fundamental $\mathbf{R}=\mathbf{56}$. The scalar manifold
has rank-$7$ and it is the real symmetric coset\footnote{%
To be more precise, it is worth mentioning that the actual relevant coset
manifold is $E_{7(7)}/[SU(8)/\mathbb{Z}_{2}]$, because spinors transform
according to the double cover of the stabilizer of the scalar manifold (see
\textit{e.g.} \cite{45,46}, and Refs. therein).} $G/H=E_{7(7)}/SU(8)$, with
dimension $70$.

\begin{table}[h!]
\begin{center}
\begin{tabular}{cccccc}
$G$ & $V$ & $n$ & isotropy~alg. & degree & interpr. $D=4$ \\[0.6cm]
$SL(2,\mathbb{C})$ & $S^{3}\mathbb{C}^{2}$ & $1$ & $0$ & $4$ & $\mathcal{N}%
=2,\mathbb{R}~(T^{3})$ \\[0.15cm]
$SL(6,\mathbb{C})$ & $\Lambda ^{3}\mathbb{C}^{6}$ & $1$ & $\mathfrak{sl}(3,%
\mathbb{C})^{\oplus 2}$ & $4$ & $%
\begin{array}{c}
\mathcal{N}=2,J_{3}^{\mathbb{C}} \\
\mathcal{N}=0,J_{3}^{\mathbb{C}_{s}} \\
\mathcal{N}=5,M_{1,2}(\mathbb{O})%
\end{array}%
$ \\[0.15cm]
$SL(7,\mathbb{C})$ & $\Lambda ^{3}\mathbb{C}^{7}$ & $1$ & $\mathfrak{g}_{2}^{%
\mathbb{C}}$ & $7$ &  \\[0.15cm]
$SL(8,\mathbb{C})$ & $\Lambda ^{3}\mathbb{C}^{8}$ & $1$ & $\mathfrak{sl}(3,%
\mathbb{C})$ & $16$ &  \\[0.15cm]
$SL(3,\mathbb{C})$ & $S^{2}\mathbb{C}^{3}$ & $2$ & $0$ & $6$ &  \\[0.3cm]
$SL(5,\mathbb{C})$ & $\Lambda ^{2}\mathbb{C}^{5}$ & $%
\begin{array}{c}
3 \\
4%
\end{array}%
$ & $%
\begin{array}{c}
\mathfrak{sl}(2,\mathbb{C}) \\
0%
\end{array}%
$ & $%
\begin{array}{c}
5 \\
10%
\end{array}%
$ &  \\[0.3cm]
$SL(6,\mathbb{C})$ & $\Lambda ^{2}\mathbb{C}^{6}$ & $2$ & $\mathfrak{sl}(2,%
\mathbb{C})^{\oplus 3}$ & $6$ &  \\[0.15cm]
$SL(3,\mathbb{C})^{\otimes 2}$ & $\mathbb{C}^{3}\otimes \mathbb{C}^{3}$ & $2$
& $\mathfrak{gl}(1,\mathbb{C})^{\oplus 2}$ & $6$ &  \\[0.15cm]
$Sp(6,\mathbb{C})$ & $\Lambda _{0}^{3}\mathbb{C}^{6}$ & $1$ & $\mathfrak{sl}%
(3,\mathbb{C})$ & $4$ & $\mathcal{N}=2,J_{3}^{\mathbb{R}}$ \\[0.3cm]
$Spin(7,\mathbb{C})$ & $\mathbb{C}^{8}$ & $%
\begin{array}{c}
1 \\
2 \\
3%
\end{array}%
$ & \multicolumn{1}{l}{$%
\begin{array}{c}
\mathfrak{g}_{2}^{\mathbb{C}} \\
\mathfrak{sl}(3,\mathbb{C})\oplus \mathfrak{so}(2,\mathbb{C}) \\
\mathfrak{sl}(2,\mathbb{C})\oplus \mathfrak{so}(3,\mathbb{C})%
\end{array}%
$} & $%
\begin{array}{c}
2 \\
2 \\
2%
\end{array}%
$ &  \\[0.3cm]
$Spin(9,\mathbb{C})$ & $\mathbb{C}^{16}$ & $1$ & $\mathfrak{spin}(7,\mathbb{C%
})$ & $2$ &  \\[0.3cm]
$Spin(10,\mathbb{C})$ & $\mathbb{C}^{16}$ & $%
\begin{array}{c}
2 \\
3%
\end{array}%
$ & $%
\begin{array}{c}
\mathfrak{g}_{2}^{\mathbb{C}}\oplus \mathfrak{sl}(2,\mathbb{C}) \\
\mathfrak{sl}(2,\mathbb{C})\oplus \mathfrak{so}(3,\mathbb{C})%
\end{array}%
$ & $%
\begin{array}{c}
2 \\
4%
\end{array}%
$ &  \\[0.3cm]
$Spin(11,\mathbb{C})$ & $\mathbb{C}^{32}$ & $1$ & $\mathfrak{sl}(5,\mathbb{C}%
)$ & $4$ &  \\[0.3cm]
$Spin(12,\mathbb{C})$ & $\mathbb{C}^{32}$ & $1$ & $\mathfrak{sl}(6,\mathbb{C}%
)$ & $4$ & $%
\begin{array}{c}
\mathcal{N}=2,6,J_{3}^{\mathbb{H}} \\
\mathcal{N}=0,J_{3}^{\mathbb{H}_{s}}%
\end{array}%
$ \\[0.3cm]
$Spin(14,\mathbb{C})$ & $\mathbb{C}^{64}$ & $1$ & $\mathfrak{g}_{2}^{\mathbb{%
C}}\oplus \mathfrak{g}_{2}^{\mathbb{C}}$ & $8$ &  \\[0.3cm]
$G_{2}^{\mathbb{C}}$ & $\mathbb{C}^{7}$ & $%
\begin{array}{c}
1 \\
2%
\end{array}%
$ & $%
\begin{array}{c}
\mathfrak{sl}(3,\mathbb{C}) \\
\mathfrak{gl}(2,\mathbb{C})%
\end{array}%
$ & $%
\begin{array}{c}
2 \\
2%
\end{array}%
$ &  \\[0.4cm]
$E_{6}^{\mathbb{C}}$ & $\mathbb{C}^{27}$ & $%
\begin{array}{c}
1 \\
2%
\end{array}%
$ & $%
\begin{array}{c}
\mathfrak{f}_{4}^{\mathbb{C}} \\
\mathfrak{so}(8,\mathbb{C})%
\end{array}%
$ & $%
\begin{array}{c}
3 \\
6%
\end{array}%
$ &  \\[0.3cm]
$E_{7}^{\mathbb{C}}$ & $\mathbb{C}^{56}$ & $1$ & $\mathfrak{e}_{6}^{\mathbb{C%
}}$ & $4$ & $%
\begin{array}{c}
\mathcal{N}=2,J_{3}^{\mathbb{O}} \\
\mathcal{N}=8,J_{3}^{\mathbb{O}_{s}}%
\end{array}%
$ \\[0.3cm]
&  &  &  &  &
\end{tabular}%
\end{center}
\caption{Non-generic, nor irregular PVS with simple $G$, of type $2$ (in the
complex ground field). To avoid discussing the finite groups appearing, the
list presents the Lie algebra of the isotropy group rather than the isotropy
group itself \protect\cite{Sato-Kimura}. The interpretation (of suitable
real, non-compact slices) in $D=4$ theories of Einstein gravity is added;
remaining cases will be investigated in a forthcoming publication}
\label{tab:fourierFg2}
\end{table}

\begin{enumerate}
\item The unique duality orbit determined by the $G$-invariant constraint $%
I_{4}>0$ is the $55$-dimensional non-symmetric coset%
\begin{equation}
\mathcal{O}_{I_{4}>0}=\frac{E_{7(7)}}{E_{6(2)}}.
\end{equation}%
By customarily assigning positive (negative) signature to non-compact
(compact) generators, the pseudo-Euclidean signature of $\mathcal{O}%
_{I_{4}>0}$ is $\left( n_{+},n_{-}\right) =\left( 30,25\right) $. In this
case, $M_{-}\left( \mathcal{Q}\right) $ given by (\ref{M-}) is the $56$%
-dimensional metric of the non-compact, non-Riemannian rigid special K\"{a}%
hler non-symmetric manifold%
\begin{equation}
\mathbf{O}_{I_{4}>0}=\frac{E_{7(7)}}{E_{6(2)}}\times \mathbb{R}^{+},
\label{uno}
\end{equation}%
with signature $\left( n_{+},n_{-}\right) =\left( 30,26\right) $, thus with
character $\chi :=n_{+}-n_{-}=4$. Through a conification procedure
(amounting to modding out\footnote{%
The signature along the $\mathbb{R}^{+}$-direction is negative \cite%
{Dualities}.} $\mathbb{C\cong }SO(2)\times SO(1,1)\cong U(1)\times \mathbb{R}%
^{+}$, one can obtain the corresponding $54$-dimensional non-compact,
non-Riemannian special K\"{a}hler symmetric manifold%
\begin{equation}
\mathbf{O}_{I_{4}>0}/\mathbb{C\cong }\widehat{\mathbf{O}}_{I_{4}>0}=\frac{%
E_{7(7)}}{E_{6(2)}\times U(1)}.
\end{equation}

\item The unique duality orbit determined by the G-invariant constraint $%
I_{4}<0$ is the $55$-dimensional non-symmetric coset%
\begin{equation}
\mathcal{O}_{I_{4}<0}=\frac{E_{7(7)}}{E_{6(6)}},
\end{equation}%
with pseudo-Euclidean signature given by $\left( n_{+},n_{-}\right) =\left(
28,27\right) $, thus with character $\chi =0$. In this case, $M_{-}\left(
\mathcal{Q}\right) $ given by (\ref{M-}) is the $56$-dimensional metric of
the non-compact, non-Riemannian rigid special pseudo-K\"{a}hler
non-symmetric manifold%
\begin{equation}
\mathbf{O}_{I_{4}<0}=\frac{E_{7(7)}}{E_{6(6)}}\times \mathbb{R}^{+},
\label{due}
\end{equation}%
with signature $\left( n_{+},n_{-}\right) =\left( 28,28\right) $. Through a
\textquotedblleft pseudo-conification" procedure (amounting to modding out $%
\mathbb{C}_{s}\mathbb{\cong }SO(1,1)\times SO(1,1)\cong \mathbb{R}^{+}\times
\mathbb{R}^{+}$, one can obtain the corresponding 54-dimensional
non-compact, non-Riemannian special pseudo-K\"{a}hler symmetric manifold%
\begin{equation}
\mathbf{O}_{I_{4}<0}/\mathbb{C}_{s}\mathbb{\cong }\widehat{\mathbf{O}}%
_{I_{4}<0}=\frac{E_{7(7)}}{E_{6(6)}\times SO(1,1)}.
\end{equation}
\end{enumerate}

(\ref{uno}) and (\ref{due}) are non-compact, real forms of $\frac{E_{7}}{%
E_{6}}\times GL(1)$, which is the type 29 in the classification of regular,
pre-homogeneous vector spaces (PVS) worked out by Sato and Kimura in \cite%
{Sato-Kimura}. From its definition, a PVS is a finite-dimensional vector
space V together with a subgroup $G$ of $GL(V)$, such that $G$ has an open
dense orbit in $V$. PVS are subdivided into two types (type $1$ and type $2$%
), according to whether there exists an homogeneous polynomial on $V$ which
is invariant under the semi-simple (reductive) part of $G$ itself. For more
details, see \textit{e.g.} \cite{Sato,Richardson,Vinberg}.

In the case of $\frac{E_{7}}{E_{6}}\times GL(1)$, $V$ is provided by the
fundamental representation space $\mathbf{R}=\mathbf{56}$ of $G=E_{7}$, and
there exists a quartic $E_{7}$-invariant polynomial $I_{4}$ (\ref{I4}) in
the $\mathbf{56}$; $H=E_{6}$ is the isotropy (stabilizer) group.\medskip

Amazingly, simple, non-degenerate groups of type $E_{7}$ (relevant to $D=4$
Einstein (super)gravities with symmetric scalar manifolds) \textit{almost}
saturate the list of irreducible PVS with unique $G$-invariant polynomial of
degree 4 (\textit{cfr.} Table 2); in particular, the parameter $n$
characterizing each PVS can be interpreted as the number of centers of the
regular solution in the (super)gravity theory with electric-magnetic duality
($U$-duality) group given by $G$. This topic will be considered in detail in
a forthcoming publication.

\section*{Acknowledgments}

My heartfelt thanks to Raja, for being so kind to invite me to participate
into such a beautiful and inspiring conference.

I would like to thank Leron Borsten, Mike J. Duff, Sergio Ferrara, Emanuele
Orazi, Mario Trigiante, and Armen Yeranyan for collaboration on the topics
covered in this review.

\end{document}